\documentclass[conference]{IEEEtran}
\def\ps@headings{%
\def\@oddhead{\mbox{}\scriptsize\rightmark \hfil \thepage}%
\def\@evenhead{\scriptsize\thepage \hfil \leftmark\mbox{}}%
\def\@oddfoot{}%
\def\@evenfoot{}}
\makeatother
\usepackage{booktabs} 
\usepackage{url}
\usepackage{svg}
\usepackage{etoolbox}
\usepackage{cuted}
\usepackage{lipsum}
\usepackage{amssymb}
\usepackage{array}
\usepackage{amsmath}
\usepackage{tikz}
\usepackage{color}
\usepackage{dirtytalk}
\usepackage{algpseudocode}
\usepackage{algorithm}
\usepackage{graphicx}
\usepackage{mdframed}
\usepackage{tcolorbox}
\usepackage{comment}
\usepackage{color, colortbl}
\usepackage{xcolor}
\usepackage{cite}
\usepackage{amsmath,amssymb,amsfonts}
\usepackage{graphicx}
\usepackage{textcomp}
\usepackage{url}
\usepackage{etoolbox}
\usepackage{cuted}
\usepackage{lipsum}
\usepackage{array}
\usepackage{color}
\usepackage{dirtytalk}
\usepackage{algpseudocode}
\usepackage{algorithm}
\definecolor{celadon}{rgb}{0.67, 0.88, 0.69}
\usepackage{listings}
\usepackage[english]{babel}
\usepackage[utf8x]{inputenc}
\usepackage{amsmath,amsthm}

\usepackage{multirow}
\usepackage{fancyvrb}

\theoremstyle{definition}

\theoremstyle{plain}

\usepackage[normalem]{ulem}

\usepackage{graphicx}
\usepackage{subcaption}
\usepackage{mwe}
\usepackage{csvsimple}
\usepackage{tabularx}
\usepackage{float}
\usepackage{arydshln}

\ifCLASSINFOpdf
\else
\fi
\begin{document}
%


\title{BFRT: \textbf{B}lockchained \textbf{F}ederated Learning for \textbf{R}eal-time \textbf{T}raffic Flow Prediction}

\author{\IEEEauthorblockN{Collin Meese\IEEEauthorrefmark{1}\IEEEauthorrefmark{2}
~~~~Hang Chen\IEEEauthorrefmark{1}\IEEEauthorrefmark{3}~~~~Syed Ali Asif\IEEEauthorrefmark{3}~~~~Wanxin Li\IEEEauthorrefmark{2}~~~~Chien-Chung Shen\IEEEauthorrefmark{3}~~~~Mark Nejad\IEEEauthorrefmark{2}}
\IEEEauthorblockA{\IEEEauthorrefmark{2}Department of Civil and Environmental Engineering, University of Delaware, U.S.A.\\\IEEEauthorrefmark{3}Department of Computer and Information Sciences, University of Delaware, U.S.A.\\
\{cmeese,chenhang,asifrabi,wanxinli,cshen,nejad\}@udel.edu~~~~\IEEEauthorrefmark{1}These authors contributed equally to this work.}
}
\vspace{-3cm}

\maketitle

\thispagestyle{plain}
\pagestyle{plain}

\begin{abstract}
Accurate real-time traffic flow prediction can be leveraged to relieve traffic congestion and associated negative impacts. The existing centralized deep learning methodologies have demonstrated high prediction accuracy, but suffer from privacy concerns due to the sensitive nature of transportation data. Moreover, the emerging literature on traffic prediction by
distributed learning approaches, including federated learning, primarily focuses on offline learning. This paper proposes BFRT, a blockchained federated learning architecture for online traffic flow prediction using real-time data and edge computing. The proposed approach provides privacy for the underlying data, while enabling decentralized model training in real-time at the Internet of Vehicles edge. We federate GRU and LSTM models and conduct extensive experiments with dynamically collected
arterial traffic data shards. We prototype the proposed permissioned blockchain network on Hyperledger Fabric and perform extensive tests using virtual machines to simulate the edge nodes. Experimental results outperform the centralized models, highlighting the feasibility of our approach for facilitating privacy-preserving and decentralized real-time traffic flow prediction.
\end{abstract}

\vspace{-3mm}

\section{Introduction}
\label{sec:introduction}

Traffic prediction plays a critical role in alleviating traffic congestion and associated negative impacts (e.g., unreliable travel time estimates, increased fuel consumption, adverse environmental effects) \cite{akhtar2021review}. Recently, as vehicle miles traveled (VMT) continues to grow annually, traffic congestion has become a pervasive societal problem. For example, the 2019 Urban Mobility Report estimated congestion in the United States resulted in an additional 8.8 billion hours of travel time and 3.3 billion gallons of extra fuel consumption in 2019 alone~\cite{schrank2019}. Consequently, real-time traffic prediction can help alleviate congestion by providing roadway users and businesses accurate real-time traffic conditions for route planning and offering rerouting options once the vehicles are en route. 

Recently, deep learning (DL) has become a promising method for traffic flow prediction (TFP), having demonstrated much success in the literature \cite{6894591}\cite{polson2017deep}, with prediction accuracy as high as 93\%. However, in the context of the TFP problem, existing DL models are centrally trained, requiring vast amounts of data to be collected and aggregated by a data center or the cloud for processing. This process makes it difficult to comply with the data privacy regulations \cite{liu2020privacy} if all of the collected data must be directly shared to facilitate model creation, or when the crowd-sourcing techniques are used \cite{zhang2020verifiable}. 
In addition, the emergence of connected and autonomous vehicles (CAVs) into the internet of vehicles (IoV) network will necessitate new methodologies, including real-time learning and prediction, for utilizing the wealth of data dynamically generated by sensors on CAVs \cite{azad2018trustvote}. 
Therefore, a paradigm shift in DL methodologies becomes necessary to enable efficient and distributed online model training leveraging real-time data while protecting data privacy.

Federated learning (FL), where multiple participants collaboratively train a learning model without exposing the underlying data \cite{bonawitz2019towards}, has been investigated as a way to enable efficient and distributed knowledge sharing for various applications~\cite{yang2019federated}. For TFP, FL can train the model in real-time and update it dynamically as traffic sensors and edge devices continuously collect incoming traffic data. 
Moreover, FL removes the need to share or aggregate locally-collected traffic data, providing improved communication efficiency, privacy, and security for stakeholders (e.g., data collectors). 

This paper proposes a novel framework for real-time traffic prediction by integrating FL with a permissioned blockchain network of edge devices. Our FL approach to real-time TFP utilizes the \emph{Federated Averaging} (\emph{FedAvg}) algorithm \cite{bonawitz2019towards} with deep learning models. Specifically, the roadside units (RSUs) collect local traffic data within their respective observance areas and then leverage it to train a localized traffic flow prediction model. 
After training, the RSUs share their knowledge with other participants by sending only the up-to-date model parameters, in contrast to sharing the local traffic data. 
In addition to preserving privacy, our approach has the added benefit of distributing training workload to the RSUs at the network edge.

We propose to use a permissioned blockchain network as the framework for FL. 
Blockchain is a decentralized network technology that can provide the benefits of reliability, security, and integrity for all stored local models in comparison to centralized storage alternatives (e.g., cloud service providers) \cite{li2020blockchain} \cite{li2021location}. 
Regarding the IoV, blockchain has been recently demonstrated as a strong candidate for improving security of the networking layer \cite{rathee2020crt}. In the case of permissioned blockchains, the consortium of participating nodes control both system usage (writes) and data access (reads) \cite{PermBlock} \cite{li2021p}. This is in stark contrast with public permissionless blockchains (e.g., Bitcoin), where anyone can join the network at will, access all of the data, and participate in consensus processing \cite{nakamoto2008bitcoin}. These properties of permissioned blockchain make it an ideal candidate for the FL framework. 

Together, permissioned blockchain and FL provides a novel way to train TFP models in real-time using locally collected live data, with additional benefits of being dynamic, privacy-preserving, decentralized and low-latency in comparison with the existing, centralized model training approaches. 
Moreover, the continued integration of newly learned model parameters from RSUs enables the global model to better react to dynamic network conditions. In contrast, existing static and centralized models suffer from a significant delay to learn from newly collected data \cite{kang2020reliable}. Notably, these benefits provide strong justification for further research into integrated blockchain and federated learning-based approaches for real-time TFP.

{\bf Contributions}.
This paper presents \emph{BFRT}, a blockchain-enabled Federated Learning approach to real-time traffic flow prediction. In this work, we make the following contributions:
\begin{itemize}
    \item We propose a privacy-preserving and secure-by-design FL framework for collaborative and real-time traffic prediction leveraging RSUs, edge devices, and permissioned blockchain. Specifically, we design federated versions of both the LSTM and GRU models and use traffic flow prediction as the example case study. 
    \item Unlike existing works \cite{liu2020privacy}\cite{chai2020hierarchical}\cite{qi2021privacy}, we design a real-time FL process where the data shards for each participant (e.g., RSU) are distinct and private. Online learning simulations are conducted to replicate real-time data collection and training of the federated model. We evaluate the learning performance of our FL models for predicting traffic in a real-world scenario using dynamically collected incoming arterial data. Experimental results show that our FL approach can generally outperform the centralized baseline models. 
    
    
    
    \item We prototype a permissioned blockchain network using Hyperledger Fabric \cite{hyperledgerfabric} and simulate the edge nodes as resource-constrained virtual machines. We evaluate the performance of the permissioned blockchain network using the Hyperledger Caliper  \cite{hyperledgercaliper} benchmarking tool. Experimental results demonstrate the blockchain network can provide suitable throughput and latency for our proposed FL architecture. 
    
  
\end{itemize}

\section{Related Work}

In recent years, FL has emerged as a novel approach for ensuring the privacy of user data in a collaborative learning setting \cite{kairouz2019advances,he2020fedml, kanagavelu2020two, 9507294}. Specifically, FL is a distributed machine learning approach that enables a model to be trained locally by a decentralized dataset hosted across various devices with different locations. 
Blockchain technology has been applied in some FL studies to achieve true decentralization during the FL model aggregation \cite{kim2019blockchained, ramanan2019baffle,ma2020federated}. Additionally, blockchain-based FL implementations leverage the blockchain features such as auditability, traceability, and immutability to make the FL approach more robust, reliable, and trustworthy \cite{bonawitz2019towards}  \cite{chen2021robust}. 

Presently, FL has been at the forefront of various collaborative learning research areas where a large amount of data is required to train the model and achieve desirable performance. The authors of \cite{elbir2020federated} discussed how FL could be used instead of centralized machine learning (ML) for building vehicular network applications in intelligent transportation systems. FL has also been implemented in several collaborative data-preserving schemes in smart AI-based IoV systems. Specifically, in the study of \cite{lim2021towards} the authors proposed an FL approach to enable privacy-preserving collaborative ML across a federation of independent Drones-as-a-Service (DaaS) providers for traffic prediction and parking occupancy management. In \cite{chai2020hierarchical} the authors simulated a hierarchical blockchain-enabled FL algorithm for knowledge sharing within the IoV to improve reliability and security.  

With the ever-growing amount of traffic data and the introduction of DL, traffic flow prediction has achieved tremendous success. However, the current prediction approaches are also concerned with the challenges regarding user privacy. Considering the performance trade-off, the authors of \cite{liu2020privacy} proposed an FL-based gated recurrent unit neural network algorithm (FedGRU) and a clustering-based ensemble scheme to organize the data-generating entities for TFP before applying FedGRU. Additionally, in \cite{zeng2021multi}, a multi-task FL framework implementing hierarchical clustering to partition collected traffic data for traffic flow prediction and route planning was proposed to cover the diverse traffic situations. Although these works have used FL and DL approaches for TFP, they have primarily evaluated their design with identical data shards using offline FL training and inference. To the best of our knowledge, no other FL studies for TFP have focused on the real-time setting for online FL learning and inference with distinct data shards. 

\begin{figure*}[ht]
\centering
\includegraphics[width=0.75\textwidth]{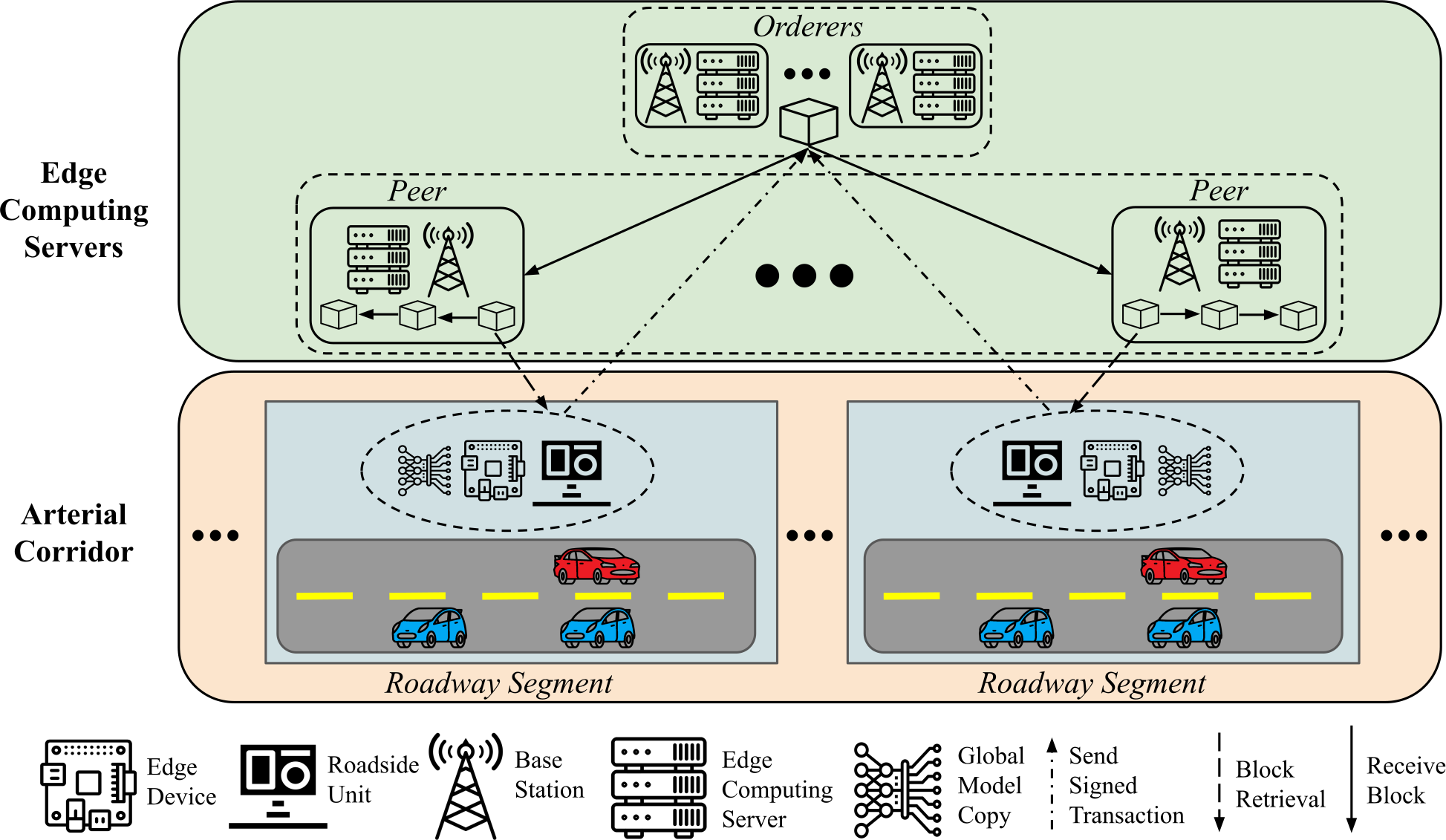}
\caption{System architecture of \emph{BFRT}.}
\label{fig:sysarch}
\end{figure*}



\section{BFRT System Design}






\subsection{Problem Definition}
In the context of our work, we define real-time FL-based TFP as a traffic prediction problem along an arterial corridor. Overseeing the corridor, there is an administration (e.g., the State's Department of Transportation) that manages and operates a set of deployed RSUs acting as traffic collection devices along the corridor. Alongside the RSUs are the edge devices that cooperatively train an online traffic prediction model leveraging the continuously collected traffic data from the RSUs in real-time, without sharing their underlying data. The edge devices also interact with the backend edge computing servers to execute the {\em FedAvg} algorithm and participate in blockchain-related operations.

\subsection{System Architecture of BFRT}
In the framework of Hyperledger Fabric, the entities of {\em clients}, {\em peers}, and {\em orderers} are defined.
Clients $\mathcal{C} = \{c_1, c_2, ..., c_{|\mathcal{C}|}\}$ are the participants who interact with the Hyperledger Fabric blockchain network but do not expend additional resources on blockchain operations (such as transaction evaluation, ordering, validation, and block creation and storage). 
Peers $\mathcal{P} = \{p_1, p_2, ..., p_{|\mathcal{P}|}\}$ represent the nodes who perform transaction execution, endorsement, and validation, as well as maintain the blockchain ledger. 
Lastly, orderers $\mathcal{O} = \{o_1, o_2, ..., o_{|\mathcal{O}|}\}$ are the nodes tasked with consensus processing that order the transactions and batch them into blocks, ensuring consistency and fault tolerance for the blockchain network. 

Fig. \ref{fig:sysarch} depicts the architectural components of {\em BFRT}, the mapping of these components to the entities of Hyperledger Fabric, and their operations and interactions. All the devices (RSUs, edge devices, and edge computing servers) are owned and operated by the administration to have the most deployment flexibility as the deployment strategy for peers and orderers can significantly impact the blockchain's performance (throughput and latency) for traffic prediction.

\textbf{RSUs} deployed along the roadway collects local traffic data in real-time. The {\bf edge device} co-located with each RSU possesses a current copy of the {\bf global traffic prediction model} and acts as a client to participate in the FL process by training the global model with its locally collected traffic data by the RSU. In {\em BFRT}, the global model can be either the Federated GRU model or the Federated LSTM model. After one round of training, each edge device submits a {\em transaction proposal} containing the updated parameters of its locally trained model, encapsulated in the HDF5 file format, to all the {\bf edge computing servers} \underline{acting as peers} for endorsements. 

In {\em BFRT}, the backend edge computing servers are deployed alongside base stations. Each peer edge computing server independently evaluates the transaction proposal and sends back an endorsement to the client. The client edge device packages all the received endorsements into a transaction, then digitally signs it, and submits it to the edge computing servers \underline{that act as orderers}. The orderers execute the \emph{RAFT} consensus algorithm \cite{ongaro2014search} to establish an unambiguous order on transactions and batch them into a new block. Note that the edge computing servers can be configured to run as either peer, orderer, or both. Each peer edge computing server receives the new block,
validates the transactions, and commits it to its blockchain ledger copy. Each client edge device downloads the block, then retrieves the encapsulated transactions, and lastly performs {\em FedAvg} on the sets of most recent model parameters to generate a new version of the global model.

\subsection{Real-time Federated Learning of Traffic Flow}
\begin{figure}[t]
\centering
\includegraphics[width=0.47\textwidth]{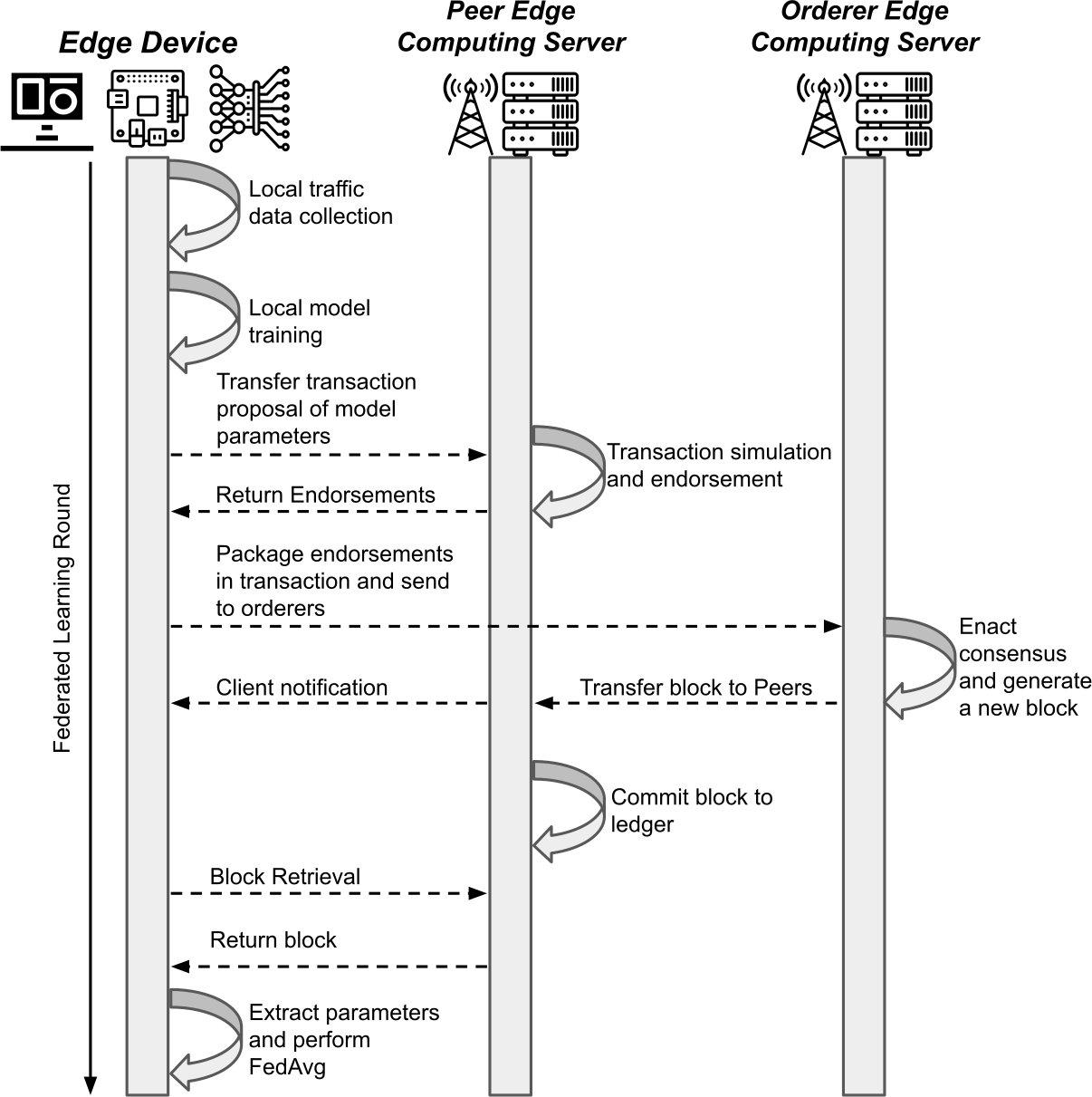}
\caption{Workflow of the \emph{BFRT} system.}
\label{fig:workflow}
\vspace*{-0.2cm}
\end{figure}

\emph{BFRT} trains a model in a real-time manner without requiring exchange or aggregation of any collected data, resulting in dynamic and efficient-to-update traffic prediction models. 
\emph{BFRT} accomplishes this by letting the clients $\mathcal{C}$ collaboratively train a single, continuously updated global prediction model~$G$ using \emph{FedAvg}, leveraging the new incoming traffic data, in a series of communication rounds $R=\langle r_1, r_2, \cdots, r_i, \cdots \rangle$. During each round $r$, every client ${c \in \mathcal{C}}$ performs the following sequence of operations: (1) collect local data; (2) train the global model with local data; (3) generate a transaction proposal with the updated model parameters; (4) send the transaction proposal to all the peers; (5) upon receiving endorsements, package the received endorsements into a transaction, digitally sign it, and submit the transaction to the orderers; (6) retrieve the newly created block; (7) extract parameters from the transactions in the block; and (8) update the global model. These operations are presented in Algo.
\ref{alg:BFRT} and Fig. \ref{fig:workflow}.

Initially, each client $c$ is provided an identical model $G^0$ before the first round $r_1$. $G^0$ could optionally be a pretrained model to jump-start the learning process. At the start of round~$r_i$, each client $c$ collects the incoming local traffic data $d^{in, i}_c$ (Algo. \ref{alg:BFRT}: line 2) for a predefined period $p$ to be combined with its local historical traffic data $d^{old, i}_c$ to form $d^{new, i}_c$ (Algo.~\ref{alg:realtime-data-collection}: lines 3-6). To mitigate overfitting the old data, data samples in $d^{new, i}_c$ should be limited to a maximum data sample size $MaxDataSize$, where data from the older rounds is excluded from the training dataset in the new rounds (Algo.~\ref{alg:realtime-data-collection}: lines 7-10). 

Following data collection, $c$ trains (updates) its local copy of the global model obtained from the last round $G^{i-1}$ using $d^{new, i}_c$ (Algo. \ref{alg:BFRT}: line 3). After that, the parameter set of the updated local model $L^i_c$ is encapsulated in the HDF5 file format and saved in a transaction proposal~$tx^i_c$ (Algo. \ref{alg:BFRT}: line 4). $tx^i_c$ is sent to all the peers $\mathcal{P}$ who evaluate and endorse~$tx^i_c$ (Algo. \ref{alg:BFRT}: lines 5-6). 
Afterwards,~$c$ packages all the endorsements into a transaction $etx^i_c$ and submits $etx^i_c$ together with its digital signature $signature_c$ to orderers $\mathcal{O}$, who reach consensus on the order of submitted transactions (Algo. \ref{alg:BFRT}: line 7) and create the new block $Block_i$ containing all the transactions. After the block $Block_i$ has been committed by peers $\mathcal{P}$ to their copy of the blockchain ledger, client~$c$ downloads $Block_i$ (Algo. \ref{alg:BFRT}: line 8) from peers, extracts the model parameter sets trained by all the client edge devices from the transactions within $Block_i$, and performs \emph{FedAvg} on the new model parameter sets to generate a new version of the global model $G^{i}$ (Algo. \ref{alg:BFRT}: lines 9-10).

\begin{algorithm}
\caption{Operations of clients $\mathcal{C}$ in round $r_i$}\label{alg:BFRT}
\begin{algorithmic}[1]
\State For each client ${c \in \mathcal{C}}$ in round $r_i$ of $R$ in parallel do:
\State $d^{new, i}_c$ $\leftarrow$ $c$.\textsc{UPDDataSet($d^{old, i}_c$)}; \Comment{Algo. \ref{alg:realtime-data-collection}}
\Statex \Comment{$d^{new, i}_c$ becomes $d^{old, i+1}_c$ in $r_{i+1}$}
\State $L^i_c$ $\leftarrow$ $c$.\textsc{TrainLocalModel($G^{i-1}$, $d^{new, i}_c$)};
\State $tx^i_c$ $\leftarrow$ $c$.\textsc{FormTransactionProposal(HDF5($L^i_c$))};
\State $c$.\textsc{SendToPeers($tx^i_c$, $\mathcal{P}$)};
\State $etx^i_c$ $\leftarrow$ $c$.\textsc{GetEndorsedTransaction($\mathcal{P}$)};
\State $c$.\textsc{SendToOrderers($etx^i_c$, , $signature_c$, $\mathcal{O}$)};
\State $Block_i$ $\leftarrow$ $c$.\textsc{BlockRetrieval($\mathcal{P}$)};
\State $\{L^i_c$, $\forall$ $c \in \mathcal{C}\}$ $\leftarrow$
\Statex \hspace{1.5cm} $c$.\textsc{ExtractParams(}\{HDF5($L^i_c$)\} in $Block_i$\textbf{)};
\State $G^i$ $\leftarrow$ $c$.\textsc{GlobalModelUPD(}$\{L^i_c$, $\forall$ $c \in \mathcal{C}\}$\textbf{)};
\Statex \Comment{by \emph{FedAvg}}
\end{algorithmic}
\end{algorithm}

\begin{algorithm}
\caption{{\textsc{UPDDataSet()}} 
 Real-time traffic data collection and training data update of client $c$ in round $r_i$}\label{alg:realtime-data-collection}
\begin{algorithmic}[1]
\State\textbf{Input:} $d^{old, i}_c$;
\State\textbf{Output:} $d^{new, i}_c$;
\While{within the data collection period $p$}
\State $d^{in, i}_c$ $\leftarrow$ $c$.\textsc{CollectData()};
\State $d^{new, i}_c$ $\leftarrow$ $d^{old, i}_c$ $\cup$ $d^{in, i}_c$;
\EndWhile
\If{$d^{new, i}_c.size > MaxDataSize$}
    \State $RemoveSize$ $\leftarrow$ $d^{new, i}_c.size - MaxDataSize$
    \State $d^{new, i}_c$.\textsc{RemoveOldData($RemoveSize$)};
\EndIf
\end{algorithmic}
\end{algorithm}

\subsection{Permissioned Blockchain Network}
\emph{BFRT} adopts a permissioned blockchain network as the FL framework. The permissioned blockchain controls and manages the FL workflow by processing the transactions $tx_c$'s, for all ${c \in \mathcal{C}}$ containing the clients' uploaded model parameters (i.e., local model updates $L_c$'s, for all ${c \in \mathcal{C}}$). Specifically, each transaction in a block consists of an HDF5 file containing the local model parameters for each RSU in the given FL round and the associated metadata (e.g., cryptographic signature). 
By storing the model updates on the blockchain, the history of FL can be retrieved and reviewed for quality assurance, preservation, auditing, and other purposes.

For the blockchain network, the data structure used for the stored model updates is as follows:
\vspace{2mm}
\begin{Verbatim}
   type ModelUpdate struct {
       FederatedID         string
       DetectorID          string
       RoundNumber         int 
       ModelParameters     HDF5
   }
\end{Verbatim}
\vspace{1mm}
The parameters within the $ModelUpdate$ data structure represent the following: $FederatedID$ refers to the name of the associated FL process (e.g., ``LSTM TFP I-95"); $DetectorID$ is the identifier for a specific client ${c \in \mathcal{C}}$ within the blockchain network; $RoundNumber$ indicates the FL round index (i.e.~$i$ in $r_i$) associated with a given HDF5 file; and lastly the $ModelParameters$ field contains the HDF5 file. 

The blockchain transaction workflow in \emph{BFRT} is modeled after Hyperledger Fabric v2.2 and separates the transaction endorsement process (execution) from the transaction ordering process (consensus). The endorsement process abides by a configurable policy specifying that a subset of peers must verify and approve the transaction before it can be ordered into blocks and committed to the chain. When a client submits a transaction, it is first sent to endorsing peers. Each peer then simulates the transaction in a containerized sandbox environment, after which the peer returns the endorsed transaction to the client. 

After collecting the necessary endorsements, the client forwards the endorsed transaction set to an orderer who enacts the consensus mechanism and packages the pending transactions into a new block. Following consensus, the new block is forwarded to all the peers who validate the transactions to verify each transaction's endorsements and metadata. Once verification is complete, the transactions are committed to the local copy of the blockchain and the client is notified. Lastly, the client retrieves the new block from a peer and executes \emph{FedAvg} to update its local version of the global model.

\section{Experimental Results}
\subsection{Experimental Design}
    \subsubsection{Setup} The \emph{BFRT} experiments are simulation-based, which were conducted on Google Colab with one NVIDIA P100 GPU, two Intel(R) Xeon(R) CPUs @ 2.30GHz, and 13.34 gigabytes of RAM. All experiments involved 7 clients for both LSTM and GRU federation, and their training samples were dynamically fed in sequence during each round to simulate real-time training. The parameters of $G^0$ are randomly initialized in every experiment instead of using a pretrained model. Each client in both LSTM and GRU federations adopts \emph{FedAvg}, with 5 local training epochs per round. 
    
    The simulations of blockchain operations were conducted on a virtual machine with 24 gigabytes of RAM and 8 cores of an Intel(R) Core(TM) i9-10900K CPU @ 3.70GHz. We simulate the proposed system of edge devices on Hyperledger Fabric version 2.2, using docker containers to represent the blockchain peers and blockchain orderers. We set resource restrictions on each container in accordance with the system architecture, providing a maximum of 2 gigabytes of RAM to each peer and 4 gigabytes of RAM to each orderer respectively while ensuring peers are provided with 50\% of the processor clock cycles given to each orderer container. In our reference deployment, we instantiate 4 peers each and 5 orderers using the \emph{RAFT} consensus algorithm. For all the blockchain experiments, we leverage the Hyperledger Caliper benchmarking tool to measure our deployment's transaction throughput and latency under various transaction loads. 

\subsubsection{Dataset} The dataset used in the experiments is from the Delaware Department of Transportation (DelDOT), which includes traffic flow data collected from DelDOT maintained roadways at a 5-minute time resolution. We select 7 nonsequential loop detectors along the I-95 north arterial to act as the FL clients in \emph{BFRT}. The selected dataset for each detector includes data ranging from the start of August 2019 until the end of September 2019\footnote{We separate 80\% of the data for real-time training and inference and save the remaining 20\% for future work involving offline prediction experiments.}.

\begin{figure*}[htp]
    \centering
    \begin{subfigure}[b]{0.475\textwidth}
        \centering
        \includegraphics[width=\textwidth]{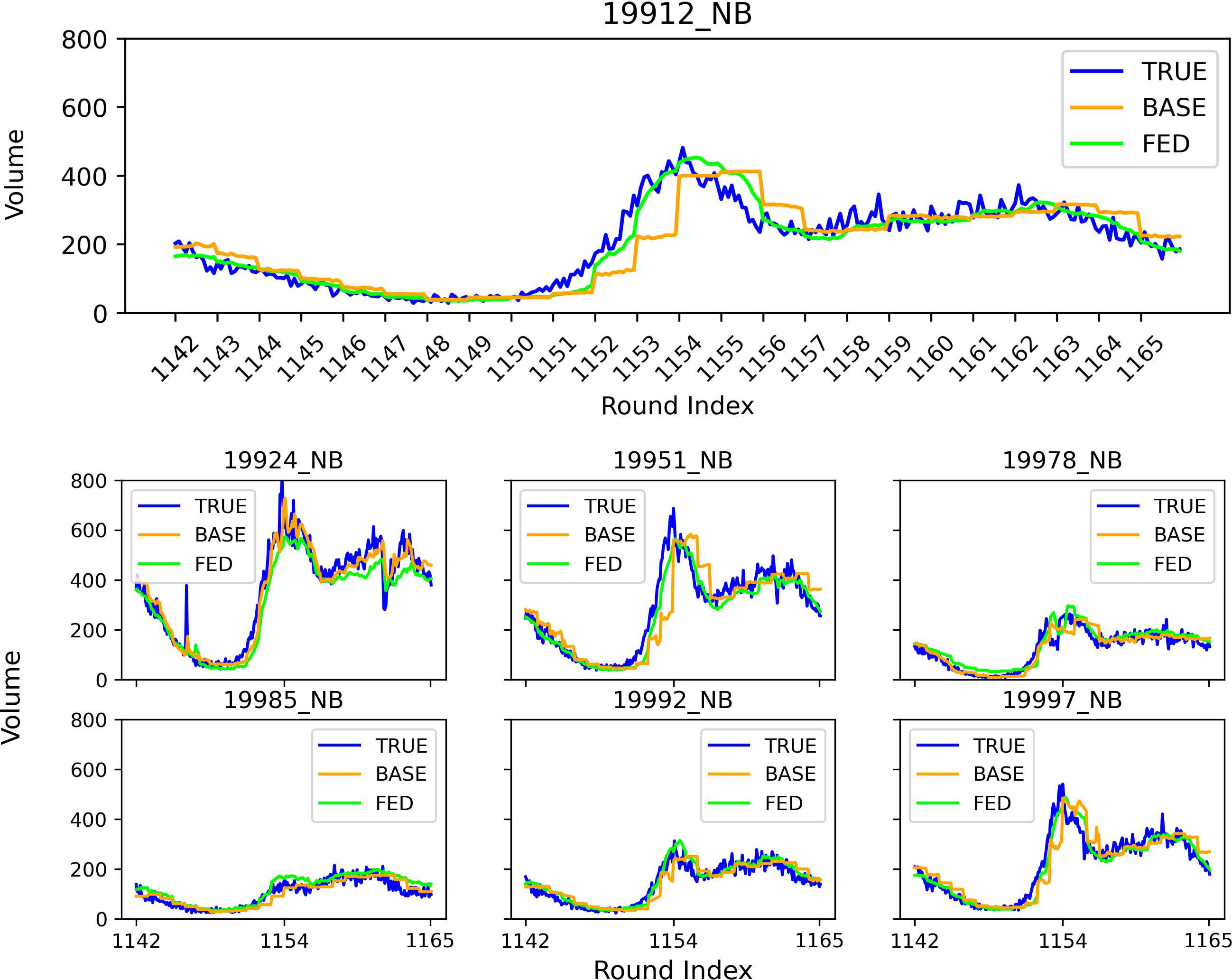}
        \caption[]%
        {{\small LSTM with $MaxDataSize = 24$}}    
        \label{fig:lstm_mdl_24}
    \end{subfigure}
    \hfill
    \begin{subfigure}[b]{0.475\textwidth}  
        \centering 
        \includegraphics[width=\textwidth]{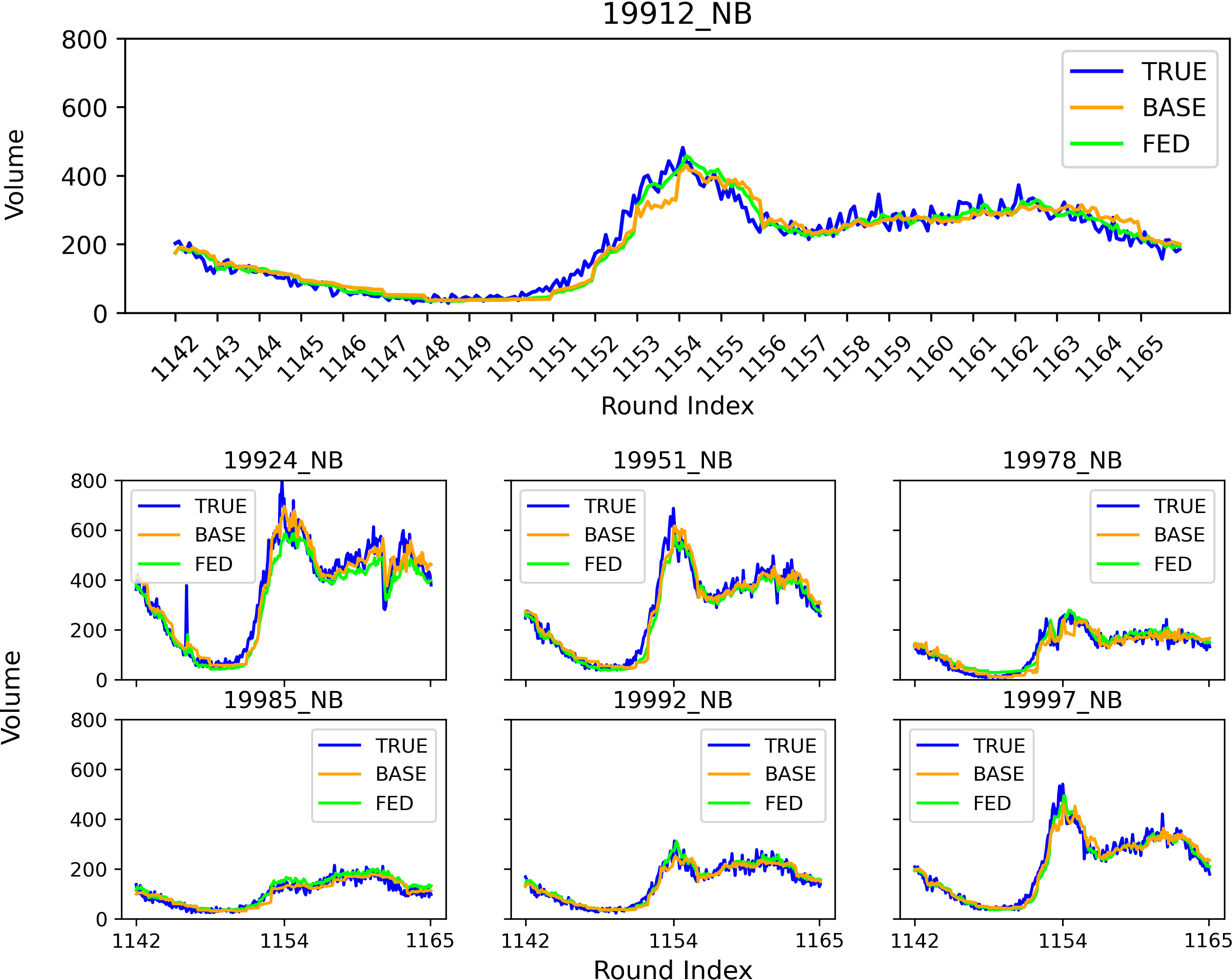}
        \caption[]%
        {{\small GRU with $MaxDataSize = 24$}}    
        \label{fig:gru_mdl_24}
    \end{subfigure}
    \vskip\baselineskip
    \begin{subfigure}[b]{0.475\textwidth}   
        \centering 
        \includegraphics[width=\textwidth]{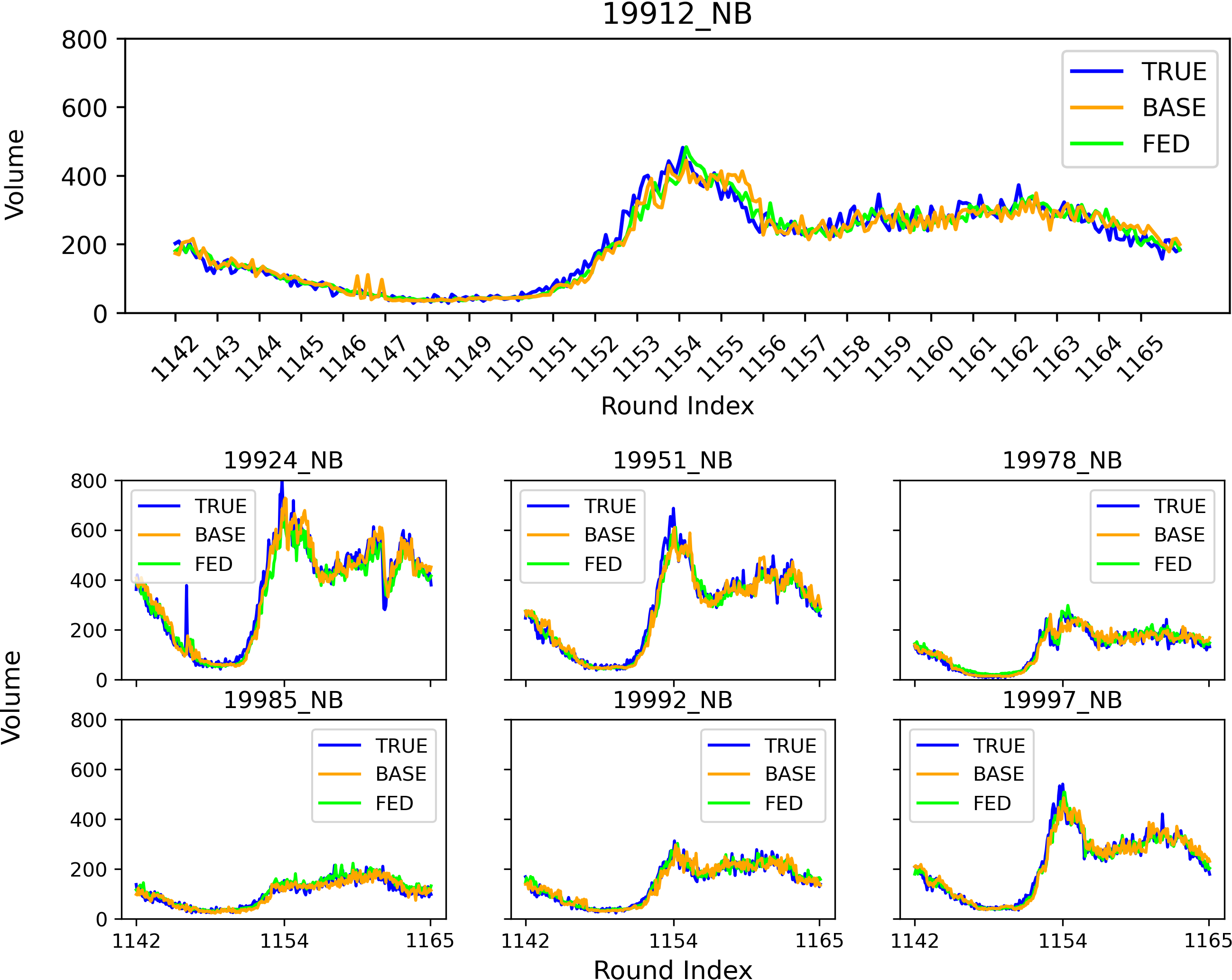}
        \caption[]%
        {{\small LSTM with $MaxDataSize = 72$}}    
        \label{fig:lstm_mdl_72}
    \end{subfigure}
    \hfill
    \begin{subfigure}[b]{0.475\textwidth}   
        \centering 
        \includegraphics[width=\textwidth]{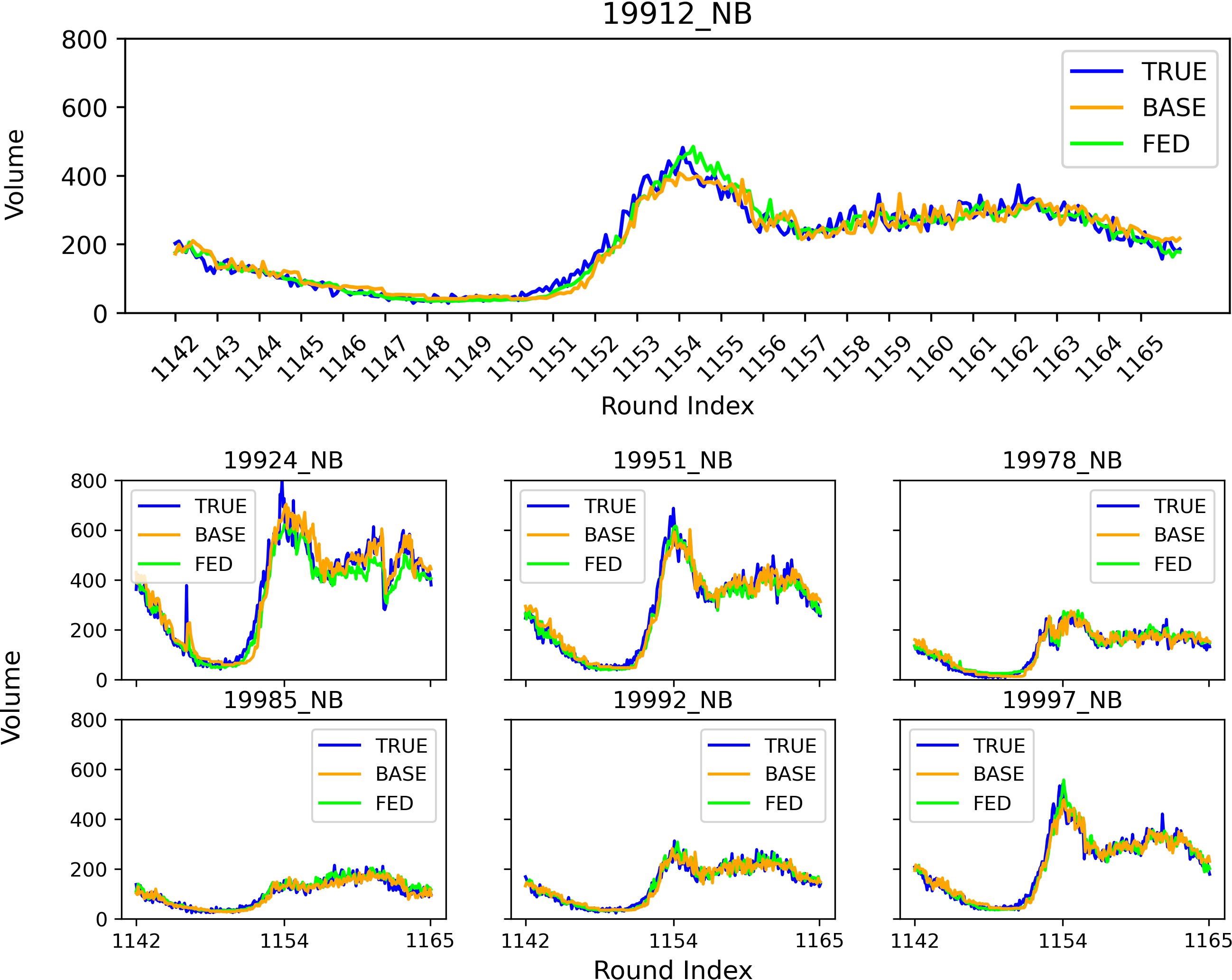}
        \caption[]%
        {{\small GRU with $MaxDataSize = 72$}}    
        \label{fig:gru_mdl_72}
    \end{subfigure}
    \caption[]
    {Real-time inferences curves of all 7 detectors during the last 24 rounds in \emph{BFRT}} 
    \label{fig:realtime_prediction}
\vspace*{-0.3cm}
\end{figure*}

\subsubsection{Models} The studies in \cite{fu2016using} found that the RNN models, specifically LSTM and GRU, exhibit comparably good inference performance on the traffic flow data from the PeMS dataset \cite{chen2002freeway} when training offline in a centralized manner. Inspired by this finding, we choose both LSTM and GRU models for real-time FL, termed \emph{LSTM-Fed} and \emph{GRU-Fed}. To highlight the performance of the federated models, we also let each client train its own centralized LSTM and GRU, termed \emph{LSTM-Base} and \emph{GRU-Base}, using the same dataset~$d^{new, i}_c$ in round $r_i$, without federation, as the baseline models. Specifically, each client $c$ in $r_i$ also uses the dynamically updated~$d^{new, i}_c$ to train either \emph{LSTM-Base} or \emph{GRU-Base} by~5 epochs as in \emph{BFRT}, but the updated baseline model $B^i$ will continue to be trained in round $r_{i+1}$, whereas in \emph{BFRT} the new global model $G^i$ will replace the local model of all ${c \in \mathcal{C}}$ when progressing to round $r_{i+1}$. The baseline training algorithm is shown in Algorithm~\ref{alg:BFRT-Base}. 

\begin{algorithm}
\caption{Baseline model training of client $c$ in round $r_i$}\label{alg:BFRT-Base}
\begin{algorithmic}[1]
\State For any client ${c \in \mathcal{C}}$ in any round $r_i \in R$:
\State $d^{new, i}_c$ $\leftarrow$ $c$.\textsc{UPDDataSet($d^{old, i}_c$)}; \Comment{Algo. \ref{alg:realtime-data-collection}} 
\State $B^i_c$ $\leftarrow$ $c$.\textsc{TrainBaselineModel($B^{i-1}_c$, $d^{new, i}_c$)};
\end{algorithmic}
\end{algorithm}

All federated and baseline RNN models in our experiments have 12 neurons in their input-layer and 1 neuron in their output-layer, respectively. This corresponds to 1 hour of traffic flow data at a 5-min resolution, which produces a 5-min look ahead prediction. As a result, the sample size of $d^{in, i}_c, i > 1$ is set to equal the $input\_shape$ of the RNN models (i.e., 12), to control data flow by hours. The exception is that $d^i_c.size, i = 1$ is set to 24 (i.e., 2 times of $d^{in, i}_c.size$, where $i > 1$) in $r_1$ for all ${c \in \mathcal{C}}$, because the models need at least 13 data samples for training. The result is a total of 1165 rounds for all \emph{BFRT} experiments. 

After designing the four models (i.e, \emph{LSTM-Base}, \emph{LSTM-Fed}, \emph{GRU-Base}, and \emph{GRU-Fed}), we perform experiments with $MaxDataSize = 24, 36, 48, 60, 72$ because the sample size is found to influence the prediction accuracy of deep learning models in \cite{ng2020influence}. In our experiments, $MaxDataSize$ controls the sample size for FL training in each round. Due to the space constraint, we report and compare the performance of the four models with $MaxDataSize = 24$ and $72$. 
Moreover, because the model architecture impacts prediction accuracy, we conducted small-scale FL simulations with varied numbers of hidden layers and neurons prior to training the LSTM models. Based on the error values, we choose 2 hidden layers and 128 neurons for each hidden layer for the LSTM model architecture. For the GRU model architecture, we make our design decisions based on the results of \cite{liu2020privacy}, selecting 2 hidden layers with 50 neurons in each. Due to the space limitations, we omit the report for the error values in these tuning experiments. To ensure reproducibility, the simulation code, selected dataset, and all other experimental results are made available in our GitHub repository\footnote{https://github.com/hanglearning/BFRT}.

\begin{figure}[ht]
\begin{subfigure}{\columnwidth}
  \includegraphics[width=\textwidth]{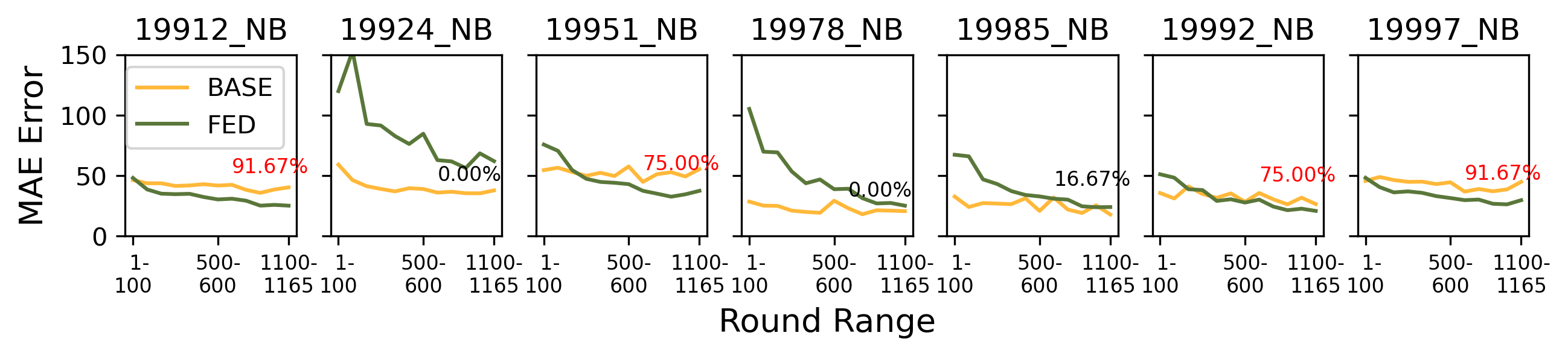}%
  \caption[]%
  {{\small MAE Error of LSTM with $MaxDataSize = 24$}}    
  \label{fig:lstm_mdl_24_error}
\end{subfigure}
\vspace{1em}

\begin{subfigure}{\columnwidth}
  \includegraphics[width=\textwidth]{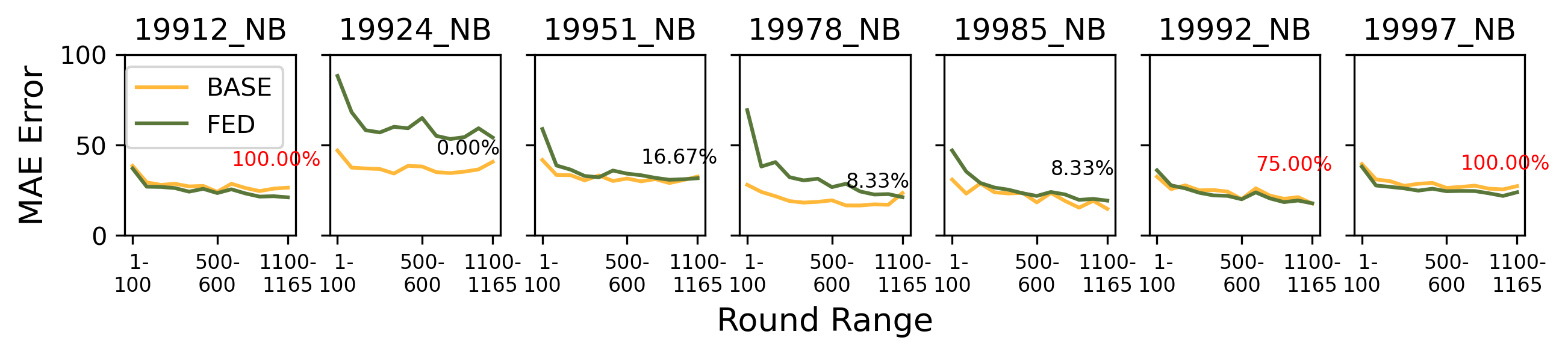}%
  \caption[]%
  {{\small MAE Error of GRU with $MaxDataSize = 24$}}    
  \label{fig:gru_mdl_24_error}
\end{subfigure}
\vspace{1em}

\begin{subfigure}{\columnwidth}
  \includegraphics[width=\textwidth]{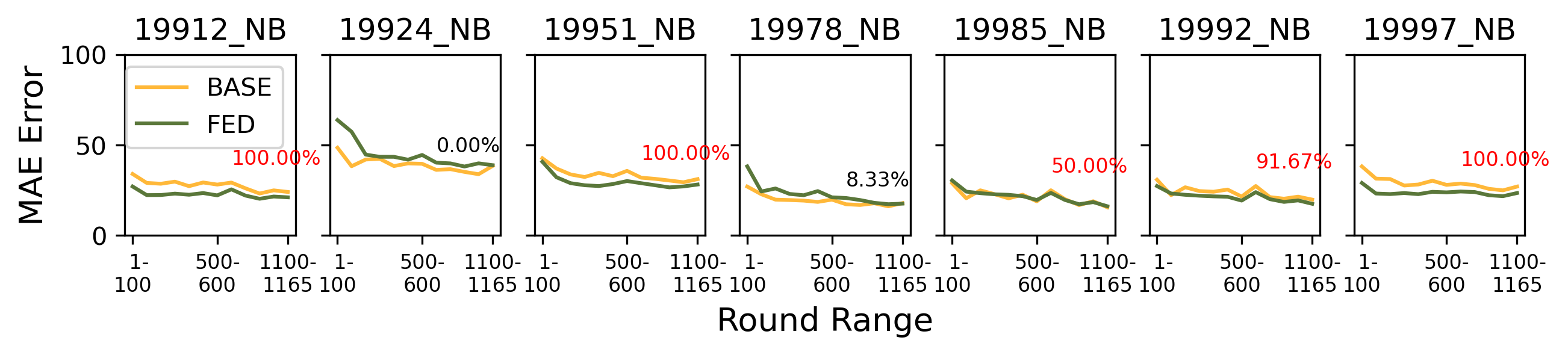}%
  \caption[]%
    {{\small MAE Error of LSTM with $MaxDataSize = 72$}}          \label{fig:lstm_mdl_72_error}
\end{subfigure}
\vspace{1em}

\begin{subfigure}{\columnwidth}
  \includegraphics[width=\textwidth]{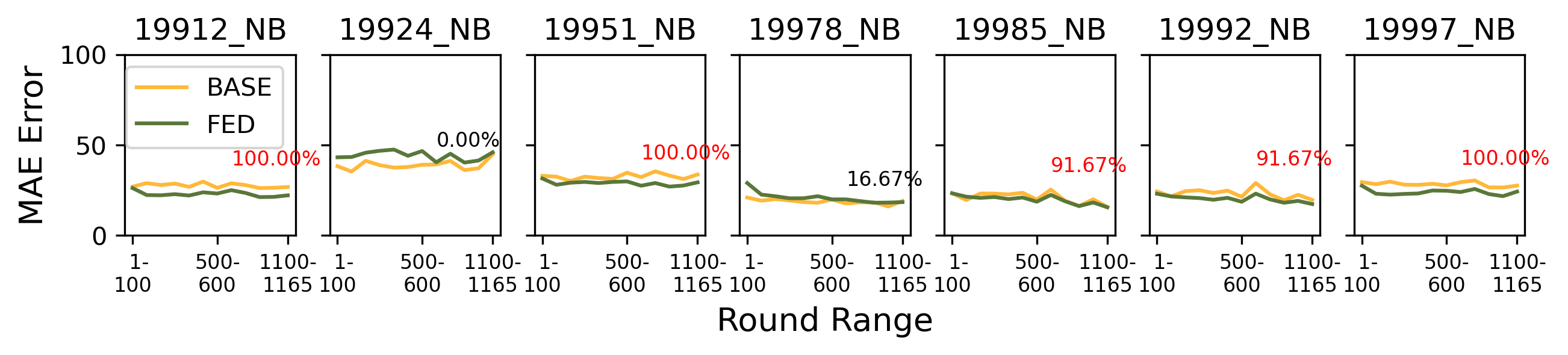}%
  \caption[]%
  {{\small MAE Error of GRU with $MaxDataSize = 72$}}      
  \label{fig:gru_mdl_72_error}
\end{subfigure}
\caption{The MAE errors of real-time inferences of all 7 detectors in all rounds of \emph{BFRT}}
\label{fig:realtime_error}
\vspace*{-0.2cm}
\end{figure}

\subsection{\emph{BFRT} Performance on Real-time Inferences}
Fig. \ref{fig:realtime_prediction} shows the partial real-time inference curves of the four models for the 7 detectors with $MaxDataSize = 24$ and $72$. Out of the inference values in the entire 1165 rounds, we report the last 24 FL rounds which includes 1 day of traffic flow data. Algo. \ref{alg:realtime_inference} describes how we obtained the inference values (i.e., the values of \emph{BASE} and \emph{FED} in Fig. \ref{fig:realtime_prediction}). 

For any client ${c \in \mathcal{C}}$ in round $r_i \in R, i > 1$, a temporary dataset $d^{pred, j}_c$ is first initialized by extracting an $input\_shape$ (i.e., 12 in our experiments, the same as $d^{in, i}_c.size, i > 1$) number of the latest data points from $d^{old, i}_c$ (Algo. \ref{alg:realtime_inference}: line 5). As a special case in $r_1$, $c$ only performs learning on $d^{in, 1}_c$, and therefore the inference starts from $r_2$. The 5-min look-ahead real-time inference in $r_i, i > 1$ is done by $c \in \mathcal{C}$ using $B^{i-1}_c$ or $G^{i-1}$ to predict on the continuously updated $d^{pred, j}_c$ (Algo. \ref{alg:realtime_inference}: lines 6-12). Specifically, the real-time inference has $input\_shape$ steps (Algo. \ref{alg:realtime_inference}: line 6), which is assumed to be exactly within the data collection period $p$. In our experiments, we set $p$ to be 1 hour to be consistent with the $input\_shape$ (i.e. 12) of the models, which in turn results in 12 steps of inference. During each step, $c$ uses $B^{i-1}_c$ or $G^{i-1}$ to predict on $d^{pred, j}_c$, and output a 5-min look-ahead prediction to be added to $BASE^{i}_c$ or $FED^{i}_c$ (Algo. \ref{alg:realtime_inference}: lines 7-8). For instance, $d^{pred, 0}_c$, which is assigned at Algo. \ref{alg:realtime_inference}: line 5, will be used to output the first prediction of the traffic volume in $r_i$. Then, $c$ waits to collect one incoming data (i.e., at a 5-min interval in our experiments) and adds it to $d^{in, i}_c$ (Algo. \ref{alg:realtime_inference}: line 9). Notably, Algo. \ref{alg:realtime_inference}: line 6, line 9 and line 10 make up the \textsc{CollectData()} function appearing in Algo. \ref{alg:realtime-data-collection}: line 4. After that, $d^{pred, j}_c$ is updated by popping out its oldest data point, and merging with $d^{in, i}_c$ (\ref{alg:realtime_inference}: line 11).

After completing an $input\_shape$ number of prediction steps, the resulted $BASE^{i}_c$ and $FED^{i}_c$ will contain the inference values from the corresponding models for $r_i, i > 1$. Lastly, $d^{in, i}_c$ is assigned to $TRUE^{i}_c$ (Algo. \ref{alg:realtime_inference}: line 13), which represents the \emph{TRUE} curve(s) in Fig. \ref{fig:realtime_prediction}.

\begin{table*}[]
\centering
\caption{Real-time Inference Errors of the Last 24 Rounds}
\label{tab:realtime-error}
\scalebox{0.87}{
\begin{tabular}{|l|l|r:r|r:r|r:r|r:r|}
\hline
DetectorID                    & Model     & MAE-24                        & MAE-72                        & MSE-24                          & MSE-72                          & RMSE-24                       & RMSE-72                       & MAPE-24                      & MAPE-72                      \\ \hline
                            & LSTM-Base & 39.68                         & 22.54                         & 3444.57                         & 992.92                          & 58.69                         & 31.51                         & 0.22                         & 0.13                         \\
                            & LSTM-Fed  & \cellcolor[HTML]{fce7af}23.07 & \cellcolor[HTML]{9AFF99}19.82 & \cellcolor[HTML]{fce7af}1076.03 & \cellcolor[HTML]{fce7af}749.4   & \cellcolor[HTML]{fce7af}32.8  & \cellcolor[HTML]{fce7af}27.38 & \cellcolor[HTML]{fce7af}0.13 & \cellcolor[HTML]{fce7af}0.11 \\ \cdashline{3-10}
                            & GRU-Base  & 24.04                         & 23.77                         & 1088.77                         & 919.01                          & 33                            & 30.32                         & 0.15                         & 0.15                         \\
\multirow{-4}{*}{19912\_NB} & GRU-Fed   & \cellcolor[HTML]{9AFF99}19.79 & \cellcolor[HTML]{fce7af}19.91 & \cellcolor[HTML]{9AFF99}717.44  & \cellcolor[HTML]{9AFF99}738.07  & \cellcolor[HTML]{9AFF99}26.79 & \cellcolor[HTML]{9AFF99}27.17 & \cellcolor[HTML]{9AFF99}0.12 & \cellcolor[HTML]{9AFF99}0.11 \\ \hline
                            & LSTM-Base & \cellcolor[HTML]{9AFF99}37.11 & 36.91                         & \cellcolor[HTML]{9AFF99}2863.12 & \cellcolor[HTML]{9AFF99}2577.12 & \cellcolor[HTML]{9AFF99}53.51 & \cellcolor[HTML]{9AFF99}50.77 & \cellcolor[HTML]{9AFF99}0.13 & 0.13                         \\
                            & LSTM-Fed  & 50.47                         & \cellcolor[HTML]{9AFF99}35.88 & 4761.38                         & 2590.1                          & 69                            & 50.89                         & 0.17                         & \cellcolor[HTML]{9AFF99}0.11 \\ \cdashline{3-10}
                            & GRU-Base  & \cellcolor[HTML]{fce7af}40.14 & 43.78                         & \cellcolor[HTML]{fce7af}2981.27 & 3441.6                          & \cellcolor[HTML]{fce7af}54.6  & 58.67                         & \cellcolor[HTML]{fce7af}0.15 & 0.17                         \\
\multirow{-4}{*}{19924\_NB} & GRU-Fed   & 45.8                          & \cellcolor[HTML]{fce7af}41.49 & 4001.79                         & \cellcolor[HTML]{fce7af}3306.1  & 63.26                         & \cellcolor[HTML]{fce7af}57.5  & 0.15                         & \cellcolor[HTML]{fce7af}0.13 \\ \hline
                            & LSTM-Base & 56.6                          & 28.39                         & 8178.46                         & 1469.3                          & 90.43                         & 38.33                         & 0.22                         & 0.12                         \\
                            & LSTM-Fed  & \cellcolor[HTML]{fce7af}33.16 & \cellcolor[HTML]{9AFF99}26.07 & \cellcolor[HTML]{fce7af}2116.6  & \cellcolor[HTML]{9AFF99}1288.38 & \cellcolor[HTML]{fce7af}46.01 & \cellcolor[HTML]{9AFF99}35.89 & \cellcolor[HTML]{fce7af}0.14 & \cellcolor[HTML]{9AFF99}0.11 \\ \cdashline{3-10}
                            & GRU-Base  & 29.19                         & 31.4                          & 1639.82                         & 1669.34                         & 40.49                         & 40.86                         & 0.14                         & 0.15                         \\
\multirow{-4}{*}{19951\_NB} & GRU-Fed   & \cellcolor[HTML]{9AFF99}28.48 & \cellcolor[HTML]{fce7af}27.84 & \cellcolor[HTML]{9AFF99}1563.93 & \cellcolor[HTML]{fce7af}1396.44 & \cellcolor[HTML]{9AFF99}39.55 & \cellcolor[HTML]{fce7af}37.37 & \cellcolor[HTML]{9AFF99}0.12 & \cellcolor[HTML]{fce7af}0.12 \\ \hline
                            & LSTM-Base & \cellcolor[HTML]{fce7af}20.91 & 16.79                         & \cellcolor[HTML]{fce7af}830.04  & 517.22                          & \cellcolor[HTML]{fce7af}28.81 & 23.9                          & \cellcolor[HTML]{9AFF99}0.21 & \cellcolor[HTML]{9AFF99}0.16 \\
                            & LSTM-Fed  & 25.79                         & \cellcolor[HTML]{fce7af}16.65 & 1039.69                         & \cellcolor[HTML]{9AFF99}466.47  & 32.24                         & \cellcolor[HTML]{9AFF99}21.6  & 0.47                         & 0.26                         \\ \cdashline{3-10}
                            & GRU-Base  & 21.81                         & 18.56                         & 877.46                          & 585.14                          & 29.62                         & 24.19                         & \cellcolor[HTML]{fce7af}0.26 & \cellcolor[HTML]{fce7af}0.24 \\
\multirow{-4}{*}{19978\_NB} & GRU-Fed   & \cellcolor[HTML]{9AFF99}20.31 & \cellcolor[HTML]{9AFF99}17.62 & \cellcolor[HTML]{9AFF99}639.12  & \cellcolor[HTML]{fce7af}515.08  & \cellcolor[HTML]{9AFF99}25.28 & \cellcolor[HTML]{fce7af}22.7  & 0.39                         & 0.29                         \\ \hline
                            & LSTM-Base & \cellcolor[HTML]{fce7af}17.11 & 15.88                         & \cellcolor[HTML]{fce7af}512.8   & 424.23                          & \cellcolor[HTML]{fce7af}22.65 & 20.6                          & \cellcolor[HTML]{fce7af}0.19 & 0.18                         \\
                            & LSTM-Fed  & 22.18                         & \cellcolor[HTML]{fce7af}15.43 & 787.55                          & \cellcolor[HTML]{fce7af}412.16  & 28.06                         & \cellcolor[HTML]{fce7af}20.3  & 0.25                         & \cellcolor[HTML]{fce7af}0.17 \\ \cdashline{3-10}
                            & GRU-Base  & \cellcolor[HTML]{9AFF99}15.07 & 14.72                         & \cellcolor[HTML]{9AFF99}376.12  & 390.9                           & \cellcolor[HTML]{9AFF99}19.39 & 19.77                         & \cellcolor[HTML]{9AFF99}0.18 & 0.16                         \\
\multirow{-4}{*}{19985\_NB} & GRU-Fed   & 17.2                          & \cellcolor[HTML]{9AFF99}14.7  & 479.97                          & \cellcolor[HTML]{9AFF99}375.67  & 21.91                         & \cellcolor[HTML]{9AFF99}19.38 & 0.2                          & \cellcolor[HTML]{9AFF99}0.16 \\ \hline
                            & LSTM-Base & 25.71                         & 20.59                         & 1155.86                         & 730.62                          & 34                            & 27.03                         & 0.22                         & 0.16                         \\
                            & LSTM-Fed  & \cellcolor[HTML]{fce7af}19.94 & \cellcolor[HTML]{fce7af}17.05 & \cellcolor[HTML]{fce7af}718.44  & \cellcolor[HTML]{fce7af}514.14  & \cellcolor[HTML]{fce7af}26.8  & \cellcolor[HTML]{fce7af}22.67 & \cellcolor[HTML]{fce7af}0.16 & \cellcolor[HTML]{fce7af}0.13 \\ \cdashline{3-10}
                            & GRU-Base  & 17.1                          & 18.77                         & 537.46                          & 613.88                          & 23.18                         & 24.78                         & 0.14                         & 0.16                         \\
\multirow{-4}{*}{19992\_NB} & GRU-Fed   & \cellcolor[HTML]{9AFF99}16.72 & \cellcolor[HTML]{9AFF99}16.69 & \cellcolor[HTML]{9AFF99}489.48  & \cellcolor[HTML]{9AFF99}494.51  & \cellcolor[HTML]{9AFF99}22.12 & \cellcolor[HTML]{9AFF99}22.24 & \cellcolor[HTML]{9AFF99}0.14 & \cellcolor[HTML]{9AFF99}0.13 \\ \hline
                            & LSTM-Base & 39.53                         & 22.91                         & 3288.69                         & 968.6                           & 57.35                         & 31.12                         & 0.2                          & 0.12                         \\
                            & LSTM-Fed  & \cellcolor[HTML]{fce7af}25.74 & \cellcolor[HTML]{9AFF99}19.76 & \cellcolor[HTML]{fce7af}1305.35 & \cellcolor[HTML]{9AFF99}788.8   & \cellcolor[HTML]{fce7af}36.13 & \cellcolor[HTML]{9AFF99}28.09 & \cellcolor[HTML]{fce7af}0.13 & \cellcolor[HTML]{fce7af}0.11 \\ \cdashline{3-10}
                            & GRU-Base  & 22.4                          & 23.89                         & 1042.65                         & 994.79                          & 32.29                         & 31.54                         & 0.12                         & 0.14                         \\
\multirow{-4}{*}{19997\_NB} & GRU-Fed   & \cellcolor[HTML]{9AFF99}19.79 & \cellcolor[HTML]{fce7af}21.44 & \cellcolor[HTML]{9AFF99}840.6   & \cellcolor[HTML]{fce7af}950.62  & \cellcolor[HTML]{9AFF99}28.99 & \cellcolor[HTML]{fce7af}30.83 & \cellcolor[HTML]{9AFF99}0.12 & \cellcolor[HTML]{9AFF99}0.11 \\ \hline
\end{tabular}
}
\end{table*}

\begin{table}[]
\centering
\caption{Best Model Count with $MaxDataSize = 24$}
\label{tab:best_model_24}
\begin{tabular}{|l|r|r|r|r|}

\hline
Model     & MAE-24                    & MSE-24                    & RMSE-24                   & MAPE-24                   \\ \hline
LSTM-Base & 1                         & 1                         & 1                         & 2                         \\ \hline
LSTM-Fed  & 0                         & 0                         & 0                         & 0                         \\ \hline
GRU-Base  & 1                         & 1                         & 1                         & 1                         \\ \hline
GRU-Fed   & \cellcolor[HTML]{9AFF99}5 & \cellcolor[HTML]{9AFF99}5 & \cellcolor[HTML]{9AFF99}5 & \cellcolor[HTML]{9AFF99}4 \\ \hline
\end{tabular}
\end{table}

\begin{table}[]
\centering
\caption{Best Model Count with $MaxDataSize = 72$}
\label{tab:best_model_72}
\begin{tabular}{|l|r|r|r|r|}
\hline
Model     & MAE-72                                         & MSE-72                                         & RMSE-72                                        & MAPE-72                                        \\ \hline
LSTM-Base & 0                                              & 1                                              & 1                                              & 1                                              \\ \hline
LSTM-Fed  & \cellcolor[HTML]{9AFF99}4 & \cellcolor[HTML]{9AFF99}3 & \cellcolor[HTML]{9AFF99}3 & 2                                              \\ \hline
GRU-Base  & 0                                              & 0                                              & 0                                              & 0                                              \\ \hline
GRU-Fed   & 3                                              & \cellcolor[HTML]{9AFF99}3 & \cellcolor[HTML]{9AFF99}3 & \cellcolor[HTML]{9AFF99}4 \\ \hline
\end{tabular}
\vspace*{-0.2cm}
\end{table}

In each subfigure of Fig. \ref{fig:realtime_prediction}, we picked the detector with the DetectorID \emph{19912\_NB} as the focus while keeping the rest of the six detectors' plots small for reference. The x-axis of all the 28 plots indicates the round index for the last 24 rounds, and the y-axis represents the traffic volume. Therefore, for instance, index 1165 on the x-axis represents $r_{1165}$, and the three curves plotted after index 1165 are $TRUE^{1165}$, $BASE^{1165}$ and $FED^{1165}$.

\subsubsection{Baseline models vs. federated models}
By comparing the proximity of the \emph{BASE} and the \emph{FED} prediction curves to the \emph{TRUE} data curve for \emph{19912\_NB} in Fig. \ref{fig:lstm_mdl_24}, we observe that \emph{LSTM-Fed} has better overall prediction accuracy than \emph{LSTM-Base} with $MaxDataSize = 24$ from $r_{1142}$ to $r_{1165}$. This trend is also seen in Table \ref{tab:realtime-error} when comparing the error values between \emph{LSTM-Base} and \emph{LSTM-Fed} for \emph{19912\_NB}. Table \ref{tab:realtime-error} denotes four types of error values (i.e. MAE, MSE, RMSE, MAPE) for analyzing predictions resulting from the four models of all 7 detectors with $MaxDataSize = 24$ and $72$ during the last 24 rounds. The number following the error type indicates $MaxDataSize$ (e.g., RMSE-72 denotes the RMSE inference error with $MaxDataSize = 72$ over the entire last 24 rounds). The values highlighted in green or beige are the smaller error values between the corresponding baseline and federated models within the relative 2-value groups. For instance, the \emph{LSTM-Fed} MAE-24 error value 23.07 of \emph{19912\_NB} is highlighted, to compare with 39.68 to show that \emph{LSTM-Fed} has smaller MAE-24 error compared to \emph{LSTM-Base} for \emph{19912\_NB}.

From Table \ref{tab:realtime-error}, it is observed that most of the inferences from the federated models outperform the baseline models. This is especially true for detectors \emph{19912\_NB}, \emph{19951\_NB}, \emph{19992\_NB}, \emph{19997\_NB}. For \emph{19985\_NB}, the baseline models always outperform the federated models with $MaxDataSize = 24$, whereas the federated models always have smaller errors than the baseline models with $MaxDataSize = 72$. This may be the result of differences in the real traffic flow trend of \emph{19985\_NB} compared to the other detectors (\emph{19912\_NB}, \emph{19951\_NB}, \emph{19992\_NB}, and \emph{19997\_NB}), due to changes in network topology (e.g., interstate exits or merges) between detector locations. 
Notably, in Table \ref{tab:realtime-error}, increasing $MaxDataSize$ results in improved accuracy for the federated model on \emph{19985\_NB}, highlighting the importance of additional historical data to rectify the difference in data trends. On the other hand, \emph{19924\_NB} and \emph{19978\_NB} have mixed results. For \emph{19978\_NB}, 9 out of the 16 pairs of error values indicate that federated models outperform its baseline models. However, For \emph{19924\_NB}, only 6 out of the 16 pairs of error values indicate that federated models outperform its baseline models. This may be the result of a non-recurrent traffic event or recurrent congestion occurring around $r_{1160}$, where a sharp drop in volume is observed.

In summary, the error values in Table \ref{tab:realtime-error} imply that most federated models outperform their corresponding baseline models in our experiments. However, as model performance is dataset dependent, we plan to perform the comparative experiments using different datasets in future work. 

\subsubsection{LSTM models vs. GRU models}
The error values highlighted in green of Table \ref{tab:realtime-error} also indicate the smallest error value among all four models for the respective detector. For instance, the \emph{GRU-Fed} MAE-24 error value 19.79 of \emph{19912\_NB} is highlighted in green, to compare with the values 39.68, 23.07 and 24.04, and highlight that \emph{GRU-Fed} has the smallest MAE-24 error compared to the three other models on detector \emph{19912\_NB}. The total counts of values highlighted in green with respect to the four models for each detector are also summarized in Table \ref{tab:best_model_24} and Table \ref{tab:best_model_72}. As seen in Table \ref{tab:best_model_24}, \emph{GRU-Fed} has the highest smallest error value count for all four types of errors across all the detectors with $MaxDataSize = 24$, whereas \emph{GRU-Fed} and \emph{LSTM-Fed} have a tie with $MaxDataSize = 72$. In conclusion, the \emph{GRU} models generally outperform the \emph{LSTM} models in our experiments.

\subsubsection{Real-time MAE errors across all 1165 rounds}
Similar to reporting error/loss values across training epochs in centralized training, we collected the real-time prediction errors to examine the model performance as the \emph{BFRT} rounds progress. Fig. \ref{fig:realtime_error} shows the normalized MAE errors in 100 round intervals for the four models with $MaxDataSize = 24$ and $72$ across all 1165 rounds for all 7 sensors. In all 28 subplots, the x-axis represents round range while the y-axis denotes the normalized MAE error for that round range. There are 12 values on the x-axis, each representing a range of 100 rounds, except the last round representing 65 rounds. For example, the y value corresponding to the x value of 500-600 represents the normalized MAE error comparing $[TRUE^{500}, TRUE^{600}]$ and $[BASE^{500}, BASE^{600}]$ (yellow line), or comparing $[TRUE^{500}, TRUE^{600}]$ and $[FED^{500}, FED^{600}]$ (green line). The percentage value in each plot indicates the percent of 100 round groups where the federated model inference error is lower than the baseline model. The percentage is colored red when the value is greater than or equal to 50\%, highlighting that the federated model outperformed the baseline model in that particular experiment.

\begin{algorithm}[t]
\caption{Real-time inference of client $c$ in round $r_i, i > 1$}\label{alg:realtime_inference}
\begin{algorithmic}[1]
\State For each client ${c \in \mathcal{C}}$ in round $r_i$ of $R, i > 1$:
\State \textbf{Input:} $G^{i-1}$, $B^{i-1}_c$, $d^{old, i}_c$; 
\State $d^{in, i}_c$, $BASE^{i}_c$, $FED^{i}_c$ $\leftarrow$ [], [], []; \Comment{Empty arrays.}
\State $j$ $\leftarrow$ 0;
\State $d^{pred, j}_c$ $\leftarrow$ $d^{old, i}_c[:input\_shape]$;
\Statex \Comment{Extract $input\_shape$ number of the latest data.}
\While{$j < input\_shape$}
    \State $BASE^{i}_c$.\textsc{Add($c$.PredictBy($B^{i-1}_c$, $d^{pred, j}_c$));}
    \State $FED^{i}_c$.\textsc{Add($c$.PredictBy($G^{i-1}$, $d^{pred, j}_c$));}
    \State $d^{in, i}_c$.\textsc{Add($c$.CollectOneData())};
    \State $j \leftarrow j + 1$;
    \State $d^{pred, j}_c$ $\leftarrow$ $d^{pred, j}_c$.\textsc{PopLeft()} $\cup$ $d^{in, i}_c$;
\EndWhile
\State $TRUE^{i}_c$ $\leftarrow$ $d^{in, i}_c$;
\end{algorithmic}
\end{algorithm}

By comparing Fig. \ref{fig:lstm_mdl_24_error} and Fig. \ref{fig:gru_mdl_24_error}, we observe that the \emph{GRU} models exhibit flatter error curves compared to the \emph{LSTM} models. This indicates that with $MaxDataSize = 24$, the \emph{GRU} models have better real-time prediction ability. Also, the \emph{FED} curves of the \emph{GRU} models for \emph{19924\_NB}, \emph{19951\_NB}, \emph{19978\_NB}, and \emph{19951\_NB} in Fig. \ref{fig:gru_mdl_24_error} have lower MAE errors in earlier rounds compared to their respective \emph{FED} curves for the \emph{LSTM} models in Fig. \ref{fig:lstm_mdl_24_error}. Additionally, when comparing Fig. \ref{fig:lstm_mdl_24_error} and Fig. \ref{fig:lstm_mdl_72_error}, and also Fig. \ref{fig:gru_mdl_24_error} and Fig. \ref{fig:gru_mdl_72_error}, we observe that increasing the $MaxDataSize$ from 24 to 72 may improve the real-time prediction accuracy in earlier rounds as the error curves quickly become smooth for all 7 sensors. Lastly, when comparing Fig. \ref{fig:lstm_mdl_72_error} and Fig. \ref{fig:gru_mdl_72_error}, we observe that both \emph{LSTM} and \emph{GRU} models have comparably good inference performance. This finding is consistent with the result in Table \ref{tab:best_model_72}. 17 out of the 28 plots have the percentage values in red, implying that the federated models outperform the corresponding baseline models in real-time prediction over the entire 1165 rounds in over 60\% of our experiments.

\begin{figure}[t]
\centering
    \begin{subfigure}{0.2\textwidth}
    \centering
    \includegraphics[width=\textwidth]{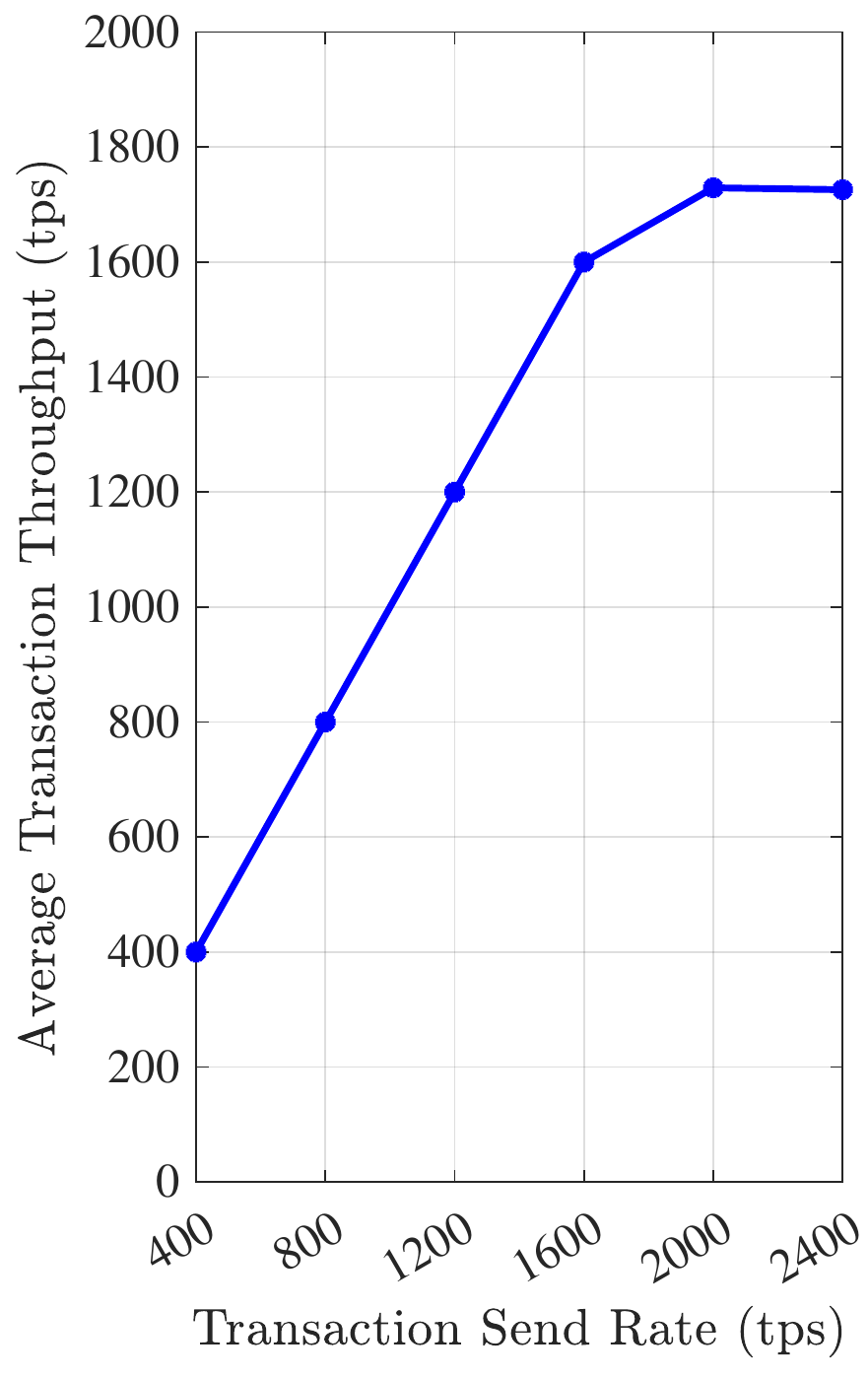}
    \caption{READ operations}
    \label{fig:READ_tps}
    \end{subfigure}
    \hfill
    \begin{subfigure}{0.204\textwidth}
    \centering
    \includegraphics[width=\textwidth]{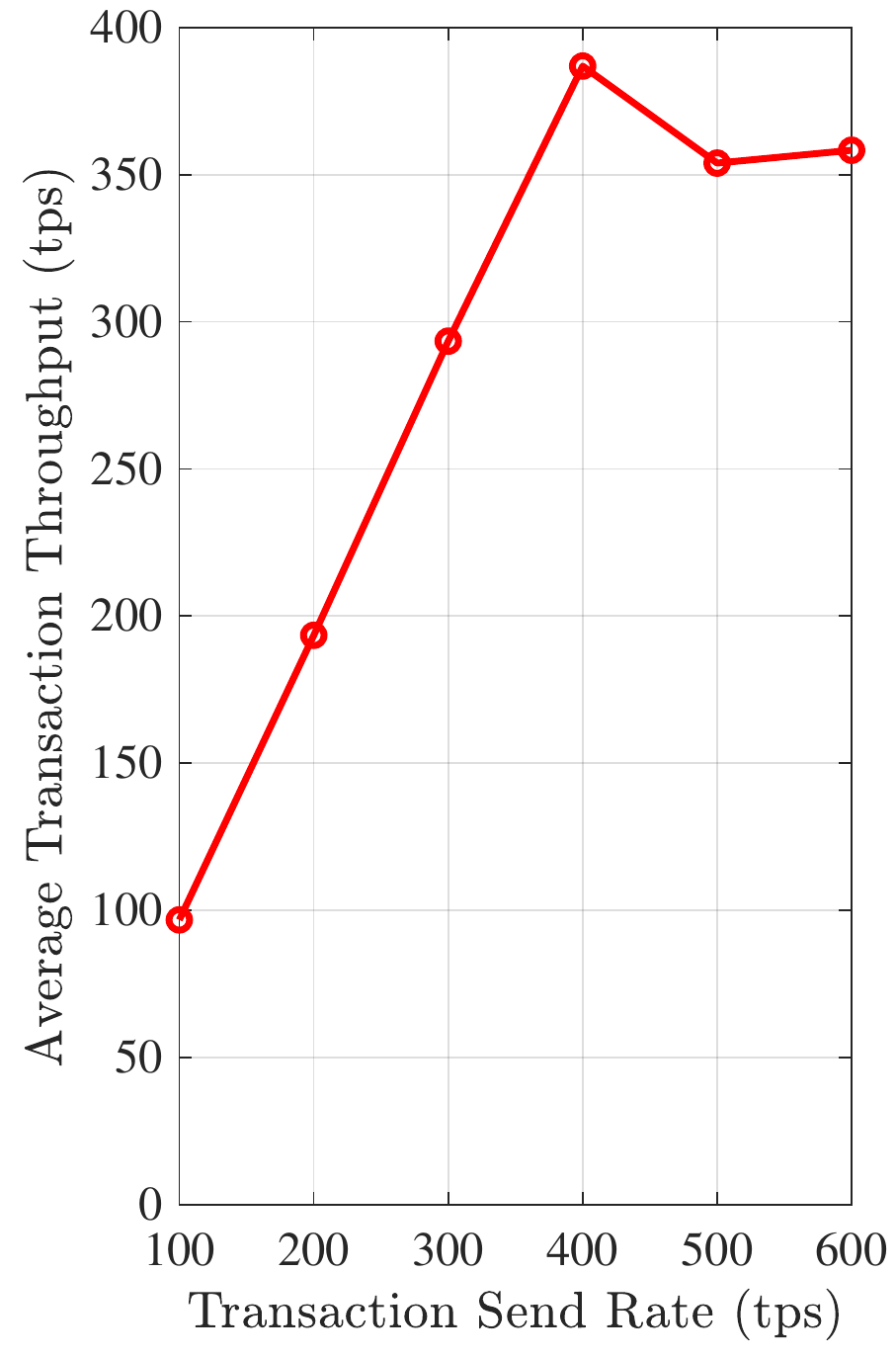}
    \caption{WRITE operations}
    \label{fig:WRITE_tps}
    \end{subfigure}
\caption{Average transaction throughputs for READ and WRITE operations.} 
\vspace*{-0.2cm}
\end{figure}

\begin{figure}[t]
\centering
\includegraphics[width=0.42\textwidth]{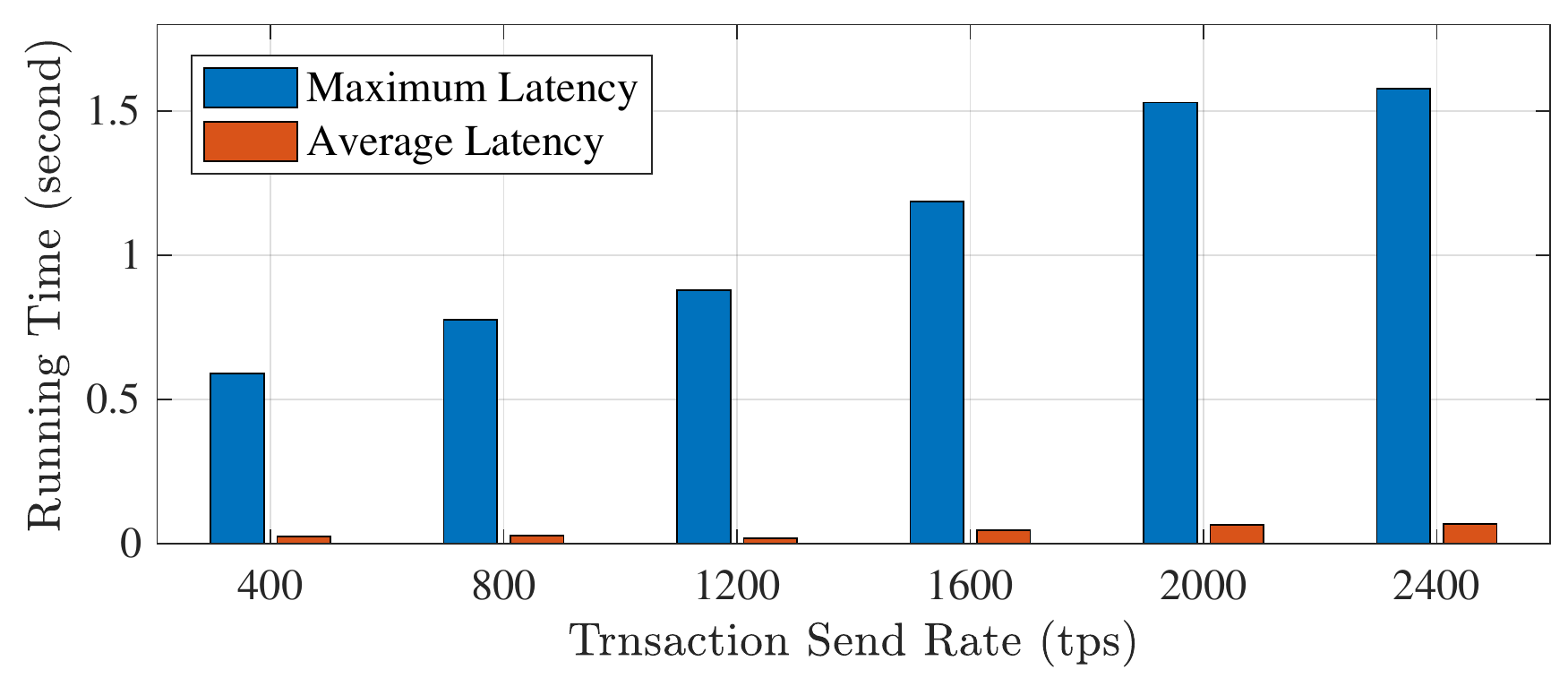}
\caption{Maximum and average transaction latencies for READ operations.}
\label{fig:READ_latency}
\vspace*{-0.2cm}
\end{figure}

\subsection{Performance of Blockchain Network}
In our blockchain experiments, we analyze the operation performance using two metrics: transaction throughput and transaction latency. Transaction throughput quantifies the number of transactions per second (TPS) which can be successfully processed by the blockchain network, while the latency indicates the running time of a single transaction from the initial construction by the client until the time it is successfully committed to the ledger. In all the experiments, we analyze both metrics under increasing send rates, where the transaction send rate indicates the number of TPS input by the blockchain clients. Notably, we choose to differentiate the operations of READ and WRITE, as the cost of both operations is not the same, and the selected operation has a notable impact on the network performance.

\begin{figure}[t]
\centering
\includegraphics[width=0.415\textwidth]{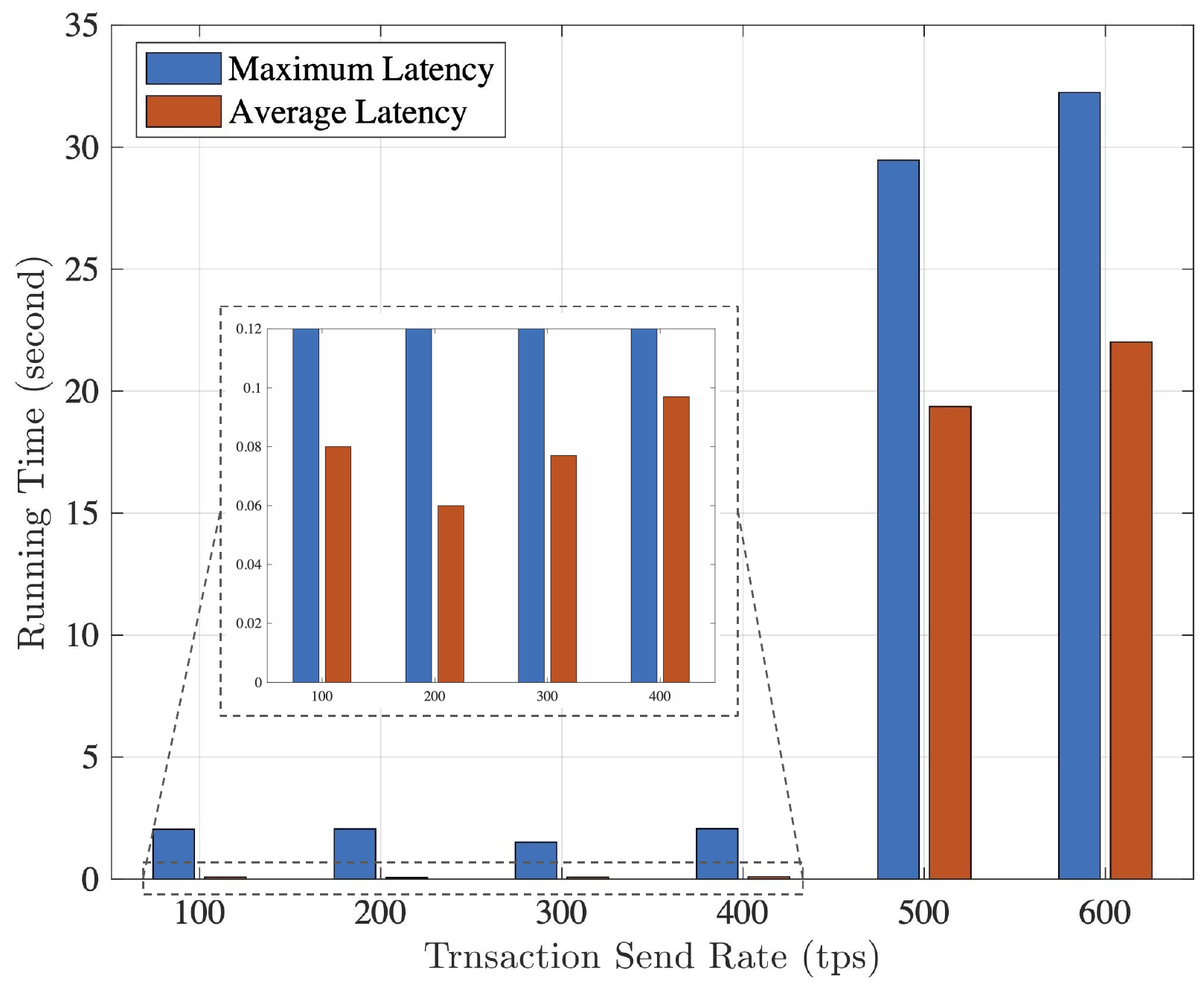}
\caption{Maximum and average transaction latencies for WRITE operations.}
\label{fig:WRITE_latency}
\vspace*{-0.2cm}
\end{figure}

\subsubsection{Transaction Throughput}
In Figs. \ref{fig:READ_tps} and \ref{fig:WRITE_tps}, we illustrate the throughput of our blockchain network under the READ and WRITE operations, respectively. We report the average transaction throughput over multiple testing cycles for quantifying the performance. Fig. \ref{fig:READ_tps} shows that the transaction throughput increases as the send rate increases to 2000. However, at send rates above 2000, the performance levels off, and throughput remains relatively constant. This indicates that the maximum network throughput for the READ operation is about 1750 TPS. 

Likewise, for the WRITE operation, Fig. \ref{fig:WRITE_tps} illustrates that the network performance begins to stabilize when the peak TPS of 390 is reached at send rate 400. These results highlight that writing is a more expensive blockchain operation than reading. Notably, the performance degrades slightly as the send rate increases beyond 400. One reason for this is the increased processing time of the WRITE operation compared to the READ operation. For the WRITE operation, as the send rate increases beyond the maximum network throughput, transactions will amass in the pending transaction pool causing a bottleneck. This issue is less visible in the READ results, as pending transactions are processed faster. 
\subsubsection{Transaction Latency}
Figs. \ref{fig:READ_latency} and \ref{fig:WRITE_latency} report both the average and minimum transaction latency for READ and WRITE operations, respectively. In Fig. \ref{fig:READ_latency} there is a clear uptrend in the latency when increasing send rate because higher transaction loads on each peer and orderer will increase confirmation times. However, it is notable that even at the highest send rate of 2400 TPS, the average latency value is still only 0.07 seconds. This demonstrates that the blockchain network can perform well for retrieval operations even under considerable transaction loads. On the other hand, Fig. \ref{fig:WRITE_latency} shows severely degraded performance when the WRITE load exceeds~400 TPS, with the average latency value rapidly converging to the maximum latency. This confirms that the blockchain will not perform well in our architecture if the WRITE load exceeds~400 TPS. Comparing both operations, it is clear that even at lower send rates, the average and maximum latency values for the WRITE operation are significantly higher than that of the READ operation.

\section{Conclusion}
This paper proposes \emph{BFRT}, a blockchained federated learning architecture for online traffic flow prediction using real-time data. \emph{BFRT} protects the privacy of underlying traffic data, while also decentralizing computation to the network edge. We prototype both the FL process and blockchain network using a combination of Python and Hyperledger Fabric, and conduct extensive experiments. In our FL experiments, we federated LSTM and GRU models and measured the real-time training and prediction performance using newly collected and distinct arterial traffic data shards. Additionally, we simulated the edge devices using a Hyperledger Fabric blockchain network with resource-limited docker containers and measured the READ and WRITE performance using the Hyperledger Caliper benchmarking tool. The results show that our FL process and models can generally outperform the centrally trained baseline models, while the permissioned blockchain network can provide high throughput and low latency. For future work, we plan to investigate new methodologies for online multi-output prediction and also experiment with various blockchain architectures to better streamline the FL workflow. We anticipate this work will lay the foundation for future research into real-time traffic flow prediction models.


\section*{Acknowledgment}
This research is supported in part by a Federal Highway Administration grant: ``Artificial Intelligence Enhanced Integrated Transportation Management System", 2020-2023. The authors highly appreciate and acknowledge the support from Gene Donaldson, DelDOT TMC's Operations Manager. 



%
\bibliographystyle{IEEEtran}
\bibliography{sig.bib}

\end{document}